\documentclass[aps,prd,amsmath,amssymb,showpacs]{revtex4}

\usepackage{epsfig,float}
\usepackage{graphicx}
\usepackage{dcolumn}
\usepackage{morefloats}
\usepackage{color}
\usepackage{slashed}
\usepackage{bbm}

\newcommand{\Lag}{\mathcal{L}}
\newcommand{\Amp}{\mathcal{A}}

\newcommand{\la}{\left\langle}
\newcommand{\ra}{\right\rangle}

\newcommand{\intlf}{\int\!\frac{\mathrm d^4 l}{(2\pi)^4}}
\newcommand{\intlt}{\int\!\frac{\mathrm d^3\vec l}{(2\pi)^3}}
\newcommand{\intlz}{\int^\infty_0\!\mathrm d l}
\newcommand{\intx}{\int^1_0\!\mathrm{d}x}

\newcommand{\mev}{~\mathrm{MeV}}

\newcommand{\gev}{~\mathrm{GeV}}

\newcommand{\dpl}{D^+}
\newcommand{\dm}{D^-}

\newcommand{\dsp}{D^{*+}}
\newcommand{\dsm}{D^{*-}}

\newcommand{\inr}{I^{(1)}}
\newcommand{\eps}{\epsilon}
\newcommand{\lef}{\left(}
\newcommand{\rig}{\right)}
\newcommand{\lt}{\left}
\newcommand{\rt}{\right}

\newcommand{\sig}{\sigma}
\newcommand{\del}{\delta}
\newcommand{\Del}{\Delta}

\newcommand{\Lam}{\Lambda}
\newcommand{\bet}{\beta}
\newcommand{\gam}{\gamma}
\newcommand{\Gam}{\Gamma}
\newcommand{\prt}{\partial}
\newcommand{\fr}{\frac}
\newcommand{\sq}{\sqrt}
\newcommand{\cd}{\cdot}
\newcommand{\rar}{\rightarrow}
\newcommand{\mr}{\mathrm}
\newcommand{\oar}{\overleftrightarrow}
\newcommand{\be}{\begin{equation}}
\newcommand{\ee}{\end{equation}}
\newcommand{\ba}{\begin{eqnarray}}
\newcommand{\ea}{\end{eqnarray}}
\newcommand{\nn}{\nonumber}
\newcommand{\ff}{\mathrm{exp}(-2\vec l^2/\Lambda^2)}
\newcommand{\ffn}{\mathrm{exp}(-2l^2/\Lambda^2)}
\newcommand{\expklm}{\mathrm{exp}(-2k^2/\Lambda^2)}

\newcommand{\fc}{\frac}
\newcommand{\st}{\sqrt}

\newcommand{\ito}{I^{(2)}_1}

\newcommand{\epstz}{\eps^{i}_{\psi}\eps^{*}_{\gam i}}
\newcommand{\epsto}{\eps_{ijk}\eps^{i}_{\psi}\eps^{*j}_\chi\eps^{*k}_\gam}
\newcommand{\epstt}{\eps^{i}_{\psi}\eps^{*j}_\gam\eps^{*}_{\chi ij}}

\begin{document}

\title{Open charm contributions to the E1 transitions of $\psi(3686)$ and $\psi(3770)\to \gamma\chi_{cJ}$ }

\author{Zheng Cao$^1$\footnote{{\it Email address:} caoz@ihep.ac.cn},
Martin Cleven$^{1,2}$\footnote{{\it Email address:} cleven@fqa.ub.edu}, Qian Wang$^{3}$\footnote{{\it Email
address:} wangqian@hiskp.uni-bonn.de}, and Qiang Zhao$^{1,4}$\footnote{{\it Email address:} zhaoq@ihep.ac.cn} }

\affiliation{$^1$ Institute of High Energy Physics and Theoretical Physics Center for Science Facilities,
        Chinese Academy of Sciences, Beijing 100049, China}

\affiliation{$^2$Departament de Fisica Quantica i Astrofisica \\
Universitat de Barcelona, 08028-Barcelona, Spain}

\affiliation{$^3$Helmholtz-Institut f\"ur Strahlen- und Kernphysik and Bethe
   Center for Theoretical Physics, \\Universit\"at Bonn,  D-53115 Bonn, Germany }

\affiliation{$^4$ Synergetic Innovation Center for Quantum Effects and Applications (SICQEA), Hunan Normal
University,Changsha 410081, China}

\begin{abstract}
The E1 transitions of $\psi(3686)$ and $\psi(3770)\to \gamma\chi_{cJ}$ are investigated in a non-relativistic
effective field theory (NREFT) where the open charm effects are included systematically as the leading corrections.
It also allows a self-consistent inclusion of the $S$-$D$ mixing in the same framework. We are able to show that the open charm contributions are essential for understanding the rather unexpected discrepancies between the
non-relativistic leading order calculations and the experimental data for these two low-lying states. 
\end{abstract}
\date{\today}

\maketitle
\section{Introduction}\label{sec:introduction}
%
Around the turn of the century the experimental possibilities of the $B$-factories together with one of their
most recognized discoveries, the mysterious $X(3872)$~\cite{Choi:2003ue}, led to a revival of charmonium
spectroscopy. A vast number of states which cannot be accommodated by the potential quark model were observed in
experiment and served as good candidates for QCD exotics such as  the charged charmonia $Z_c(3900)$ and
$Z_c(4020/4025)$ at BESIII~\cite{Ablikim:2013mio,Ablikim:2013emm,Ablikim:2013wzq,Ablikim:2013xfr} and
$Z_c(4430)$ at Belle~\cite{Choi:2007wga} and LHCb~\cite{Aaij:2014jqa}. While their nature has not been
unambiguously determined, it appears obvious that the proximity of open charm thresholds, i.e. $D^*\bar D +c.c.$
for $X(3872)$ and  $Z_c(3900)$, $D^*\bar D^*$ for $Z_c(4020/4025)$, must be closely related to their formation.
Given that these may be the outstanding examples for the importance for open thresholds, there should be other
cases that the open thresholds play a crucial role in understanding some of the open questions even for
low-lying states. As studied in Ref.~\cite{Zhang:2009kr}, the non-$D\bar{D}$ decay of $\psi(3770)$ can be
strongly affected by the intermediated $D$-meson loops via the rescattering process. This turns out to be a
natural explanation for this decay and has implications on various processes that can be tested explicitly in
experiment. In Refs.~\cite{Eichten:1979ms,Eichten:1978tg} the open threshold effects on the spectrum were
partially considered.

In the charmonium region, electromagnetic (EM) transitions serve as a crucial probe of hadron structures and
help to establish the constituent degrees of freedom within hadrons.  Within the EM transitions between
charmonium states, E1 transitions have been better measured in experiment due to their relatively enhanced
couplings with respect to the magnetic transitions because the latter are relatively suppressed by a factor of
$p_Q/M_Q$ with $p_Q$ and $M_Q$ denoting the momentum and mass of the heavy quark, respectively. In the framework
of non-relativistic quark model many theoretical studies of the heavy quarkonium EM transitions have been
carried out. For instance, the Cornell potential model~\cite{Eichten:1979ms} has been a great success in the
description of the charmonium spectrum with a spin-independent color Coulomb plus linear scalar potential. While
this is an indication of the approximate heavy quark spin symmetry (HQSS) within the charmonium system, one can
also observe deviations due to the HQSS breaking. One source for the HQSS breaking is the spin-dependent
interaction which will introduce relativistic corrections to the quark
potentials~\cite{Godfrey:1985xj,Barnes:2005pb}. A detailed review of different approaches for the charmonium EM
transitions can be found in Ref.~\cite{Brambilla:2004wf} and references therein.

In this work, we study the open charm effects on the E1 transition of $\psi(3686)$ and $\psi(3770)$ (denoted by
$\psi'$ and $\psi''$, respectively, in the following for simplicity) to $\gamma\chi_{cJ}$ in a non-relativistic
effective field theory (NREFT).  These effects from the intermediate meson loops will introduce the main
corrections to the leading NREFT results in the same framework as a natural dynamic mechanism for breaking the
HQSS. Such corrections, contributing at the order of $v^2$ in the transition amplitude with $v$ denoting the typical non-relativistic velocity of the intermediate charmed mesons, can be regarded as relativistic corrections to the charm quark potential~\cite{Eichten:1979ms}.

Since $\psi'$ and $\psi''$ both are close to the mass threshold of $D\bar{D}$, it is natural to expect that the
large couplings of these two states to the $D\bar{D}$ channel will allow us to recognize the open charm effects
and investigate their impact on the decay modes of these two states. Also, the small momentum carried by the
intermediate charmed mesons allows for the application of the NREFT to the heavy meson loops. It should be noted
that since the couplings for $\psi'$ and $\psi''$ to the open charm channels are via $P$-wave, the self-energy
corrections from the charmed meson loops are expected to be small and they can be absorbed into the physical
masses adopted in the calculation. However, the threshold effects may still have significant impact on their
decays~\cite{Zhang:2009kr}. This may appear in exclusive decays and the E1 transitions of $\psi', \ \psi''\to
\gamma\chi_{cJ}$ are ideal for probing this mechanism.

In this work we will describe the radiative E1 transitions of $\psi'$ and $\psi''$ by a leading contact
interaction that obeys the HQSS. This term will mimic the leading order results from the non-relativistic quark
model calculations. Then, three subleading contributions will be introduced by the open charm effects. First,
due to the proximity to each other and to the $D\bar D$ threshold  these two vector states can arise from mixing
of the quark model states $\psi(2S)$ and $\psi(1D)$  via charmed meson loops. Secondly, the photon can arise
from the couplings of $\psi'$ or $\psi''$ to $D\bar D$ when gauging the derivative term in the couplings where
$D\bar D$ couple to $\chi_{cJ}$. Thirdly, the transitions can be mediated by intermediate triangle $D$-meson
loops where the photon will be radiated by intermediate $D$-mesons. All four contributions can be included
self-consistently in the NREFT.

In the following we first present the NREFT framework  in Sec.~\ref{sec:framework}. The  results are discussed
in Sec.~\ref{sc:re} and a brief summary is  given in the last section.

\section{Framework}\label{sec:framework}
The E1 transitions of $\psi'$ and $\psi''$ to the leading meson loop corrections can be illustrated by
Fig.~\ref{fig:nlo}. The tree-level diagram of Fig.~\ref{fig:nlo} (a) represents the leading E1 transition
amplitude that can be compared with the potential quark model calculations. The open charm effects can
contribute as corrections to the leading tree-level amplitude via either the state mixing (Fig.~\ref{fig:nlo}
(b)), the term from gauging the couplings of $\psi'$ or $\psi''$ to $D\bar D$ (Fig.~\ref{fig:nlo} (c))  or the
intermediate triangle meson loop transitions (Fig.~\ref{fig:nlo} (d)).

In this section we will introduce the interaction Lagrangians necessary to describe the leading and
next-to-leading order processes for the E1 transitions of $\psi'$ and $\psi''$. We begin with the initial $S$-
and $D$-wave charmonia which are given by~\cite{Guo:2013zbw}
\begin{equation}
 J=\vec\psi_S\cd\vec\sig,  \qquad
 J^{ij}=\fr{1}{2}\sq{\fr{3}{5}}\lef\sig^i\psi_D^j+\sig^j\psi_D^i\rig-\fr{1}{\sq{15}}\del^{ij}\vec\sig\cd\vec\psi _D,
\end{equation}
where $\psi_S$ and $\psi_D$ annihilate $\psi'$ and $\psi''$, respectively. Note that for simplicity we have
omitted the spin partners $\eta_c$ and $\eta_{c2}$ which are irrelevant to this work. The same holds for the
$h_c$ in the case of the $P$-wave charmonia which are collected in the following multiplet~\cite{Guo:2010ak}:
\be
  \chi^i=\sig^j\lef-\chi^{ij}_{c2}-\fr{1}{\sq2}\eps^{ijk}\chi^k_{c1}+\fr{1}{\sq3}\del^{ij}\chi_{c0}\rig.
\ee
As mentioned before the leading contributions to the E1 transitions are given by the contact interactions \ba
 \Lag_{SP\gam}= g_{SP\gam}\la \chi^{\dag i}J\ra E^i+h.c., \qquad
 \Lag_{DP\gam}= g_{DP\gam}\la \chi^{\dag i}J^{ij}\ra E^j+h.c.,  \label{eq:Contact}
\ea
where we leave the couplings $g_{SP\gam}$ and $g_{DP\gam}$ to be determined from experiment.

To study the subleading contributions via charmed meson loops we need to introduce heavy meson multiplet
consisting of pseudoscalar $P_a$ and vector $V_a$ mesons. The corresponding fields for the charmed ($H_a$) and
anti-charmed meson ($\bar H_a$) are written as the following, respectively,
\be
 H_a=\vec V_a\cd\vec\sig+P_a, \qquad \bar H_a = -\vec{\bar V}_a\cd\vec\sig+\bar P_a,
\ee
where $a$ is the SU(3) flavor index and $P_a(V_a)\equiv (D^{(*)0},D^{(*)+},D^{(*)+}_s)$. These fields and their
interactions have been studied in detail in Refs.~\cite{Colangelo:2003sa,Guo:2010ak,Guo:2013zbw}.

We start with the $S$-wave charmonium $\psi'$. The coupling to a pair of charmed mesons is in a relative
$P$-wave and given by
\ba \label{eq:g2}
  \Lag_{HH\psi'}&=&i\fr{g_2}{2}\la \bar H_a^\dag\sig^i\oar\prt^i H_a^\dag J\ra+ h.c.,
\ea
where $A\oar\prt B\equiv A(\vec\prt B)-(\vec\prt A)B$. The coupling constant $g_2$ can not be determined
directly in experiment so we adopt it from Ref.~\cite{Chen:2012qq}. There, it was determined by a fit of the
lineshape of $e^+e^-\rar D\bar D$ at the mass of $\psi''$. It was shown that the interference from the $\psi'$
accounted for the anomalous lineshape and provided a reliable constraint on the coupling constant $g_2$. Note
that because of the non-relativistic normalization in the NREFT our coupling differs by a factor
$m_D\sq{m_{\psi'}}$ and reads $g_2=(-1.90\pm1.09)~\gev^{-3/2}$.

The coupling for the first $D$-wave charmonium $\psi''$ to $D\bar{D}$ is given by
\ba\label{eq:g3}
  \Lag_{HH\psi''}&=&i\fr{g_3}{2}\la \bar H_a^\dag\sig^i\oar\prt^j H_a^\dag J^{ij}\ra+ h.c. ,
\ea
where the coupling constant can be easily extracted from experiment. Using the value quoted in
PDG~\cite{Agashe:2014kda}, ${\Gamma_\mathrm {Exp}(\psi''\rar D\bar D)=\lef25.3\pm2.9\rig\mev}$, we can determine
$g_3=(2.80\pm0.15)~\mr{GeV^{-3/2}}$. It should be noted that the couplings $g_2$ and $g_3$ are extracted from experimental data. Thus, they have already been the ``dressed" couplings. Since the vertices involve $P$-wave interactions and the loop corrections are perturbative in comparison with the tree amplitudes, the vertex renormalization is expected to be insignificant. For the one loop contributions to the amplitudes it is acceptable to adopt those extracted values for $g_2$ and $g_3$ as the leading approximation. This will also reduce the number of free parameters in the formulation.

Finally, we need to consider the coupling of the $\chi_{cJ}$ multiplet to a pair of charmed mesons:
\ba\label{eq:g1}
  \Lag_{HH\chi}&=&i\fc{g_1}{2}\la \chi^{\dag i}H_a\sig^i \bar H_a\ra+ h.c.,
\ea where the parameter $g_1$ will be determined by the numerical fit. The QCD sum rule analysis has given a
prediction in Ref.~\cite{Colangelo:2002mj} which is $g_1=-4.18~\mathrm{GeV}^{-1/2}$.


In order to calculate the diagrams depicted in Fig.~\ref{fig:nlo}(d), we still need to describe the photon
coupling to the charmed meson pair. The corresponding electronic and magnetic Lagrangian for the photon coupling
to the $S$-wave charmed mesons reads~\cite{Hu:2005gf}
\ba \Lag_{HH\gam e}&=&\fc{ie}{2m_H}\la
H^\dag_a\oar\prt^iH_b(Q_H){ab}\ra A^i+h.c.,
\\\Lag_{HH\gam m}&=&\fc{e\bet}{2}\la H^\dag_aH_b\vec\sig\cd\vec BQ_{ab}\ra
  +\fc{eQ'}{2m_Q}\la H^\dag_a\vec\sig\cd\vec BH_a\ra+ h.c.,\label{eq:Lgam}
\ea
where $Q_H$ is the matrix containing the charge fractions of the heavy mesons,
$B^i=\eps^{ijk}\prt^jA^k$ is the magnetic field and $e$ is the electric charge.
The light and heavy quark charge fractions are given by
$Q=\mathrm{diag}\lef2/3,-1/3,-1/3\rig$ and $Q'=2/3$, respectively. The parameter $\beta$ can be related to the light
constituent quark mass via $\bet=1/m_q$. Further, we adopt the values of the light and heavy quark masses
$m_q=356~\mr{MeV}$ and $m_Q=1.5~\mr{GeV}$ from Ref.~\cite{Hu:2005gf} in the calculation.

\section{Results and discussion}\label{sc:re}
\begin{figure}[t] \vspace{0.cm}
\begin{center}
\hspace{20cm}
\includegraphics[scale=0.5]{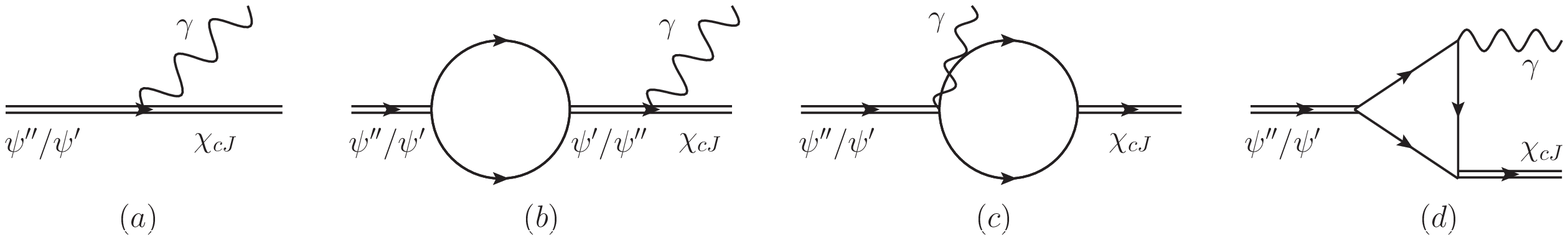}
\caption{The E1 transitions via (a) leading tree-level diagram, (b) $2S-1D$ mixing, (c) EM gauging term and (d)
intermediate meson loops. } \label{fig:nlo}
\end{center}
\end{figure}
Using the Lagrangians and couplings introduced in the  previous section, we are now ready to calculate the
partial decay widths for the six channels $\psi',\psi''\rar\gam\chi_{cJ}$. The possible processes are shown in
Fig.~\ref{fig:nlo}. Each channel includes the tree-level E1 transition, state mixing between $\psi'$ and
$\psi''$, electromagnetic (EM) contact term, and the meson loop transition
via the triangle loops, where the last two terms are given by gauging the charged strong coupling via the EM minimal substitution. The total transition amplitude can be expressed as
\begin{equation}\label{total-amp}
{\cal M}\equiv {\cal M}_{tree}+[{\cal M}_{mixing}+{\cal M}_{gauging}+{\cal M}_{triangle}]e^{i \delta} \ ,
\end{equation}
where the amplitude ${\cal M}_{tree}$ can be extracted directly from the Lagrangian in Eq.~(\ref{eq:Contact})
while ${\cal M}_{mixing}$, ${\cal M}_{gauging}$and ${\cal M}_{triangle}$ will be given by the mixing, contact
gauging term and triangle loops, respectively. A relative phase factor $\exp(i\delta)$ between the tree and loop
amplitudes seems to be necessary here. It can be interpreted as hadronic effects arising from the fact that the hadrons are not point-like fundamental particles and it also indicates the breaking of the HQSS in the charmonium system. Therefore, at least one phase angle can be introduced between the tree and loop amplitudes as a free parameter. In fact, if we require that these three loop transitions share the same phase angle with respect to the tree amplitude, the relative signs between the coupling constants will be fixed.


For the mixing amplitudes of Fig.~\ref{fig:nlo} (b) and the EM gauging amplitudes of Fig.~\ref{fig:nlo} (c) the
loop integral is divergent. For the triangle amplitudes with electronic photon couplings to the $S$-wave charmed
mesons, the loop integral is also divergent because of the two $P$-wave couplings of the $\psi$ and photon.
Thus, we introduce an overall exponential form factor
\be \label{eq:ff}
f_\Lambda(\vec l)=\mathrm{exp}(-2\vec l\,^2/\Lambda^2) \ ,
\ee
where $\vec l$ denotes the three-momentum of the charmed meson in the center-of-mass (c.m.) frame of the
charmonium system. This form factor arises from the typical two-body quark wave function convolutions in meson radiative transitions and the effective range of $\Lambda$ about 1 GeV corresponds to the typical size of hadrons. In the numerical calculations we take a range of $\Lambda=0.8\sim 1.2$ GeV as a test of the sensitivity of the loop corrections to the cut-off energy. 
The detailed calculation of the loop integrals can be found in App.~\ref{app:mixing}.

For the charmed meson loops we will consider the SU(2) and SU(3) flavor symmetry for the light quark degrees of
freedom independently to study the effects of including charmed-strange mesons in the calculation. This means in
the SU(2) scheme the charmed mesons $D^{0(*)}$ and $D^{+(*)}$ and the charge conjugations of them are allowed as
the intermediate mesons while for the SU(3) the charmed-strange meson $D_s^{+(*)}$ and their charge conjugations
will be included. The amplitudes for all channels are listed in detail in App.~\ref{app:amplitudes}.


The calculation leaves us with a total of four free parameters, namely, two coupling constants from the leading
E1 transition defined in Eq.~(\ref{eq:Contact}), the coupling of the $\chi_{cJ}$ multiplet to a pair of charmed
mesons $g_1$, and the phase angle $\delta$ defined in Eq.~(\ref{total-amp}). These parameters will be determined
by fitting the experimental data for $\psi'$ and $\psi''\to
\gamma\chi_{cJ}$~\cite{Ablikim:2015sol,Agashe:2014kda,BESIII:2015cby}.

We found several reasonable fits with form factor parameter $\Lambda$ within the region of 0.8-1.2 GeV for
channels of $\psi''\rar\gamma\chi_{c0,1,2}$ and $\psi'\rar\gamma\chi_{c0,1}$. We cannot include the channel of
$\psi'\rar\gamma\chi_{c2}$ to obtain an improved fit. Since the data are poor for this channel we expect that
more precise measurement of this channel will provide a test of our scenario in the future. In this sense our
results for this channel can be regarded as a rough prediction. The results for the best fits in these two
approaches are listed in Table~\ref{tab:para} with the reduced $\chi^2$ varying in the range of 0.66-1.95 for
different $\Lambda$ values. We should note that the couplings for $\psi'$ and $\psi''$ to $D\bar{D}$, i.e. $g_2$
and $g_3$, have opposite signs which is well established by the study of the cross section lineshapes of
$e^+e^-\to D\bar{D}$~\cite{Chen:2012qq,Zhang:2009gy}. If we further require that these two transitions share the
same phase angle $\delta$ the best fitting turns out to favor all positive sign for the tree-level couplings and
the $\chi_{cJ}$ multiplet to $D\bar{D}$ coupling $g_1$. Actually, these three fitted couplings could also be all
negative to meet the result of $g_1$ in Ref.~\cite{Colangelo:2002mj}. It will only add an overall negative sign
to all amplitudes. In both SU(2) and SU(3) schemes it shows destructive interferences between the loop and tree
amplitudes in $\psi''\rar\gamma\chi_{c0/1/2}$ channels and constructive interferences in
$\psi'\rar\gamma\chi_{c0/1/2}$ which will be discussed later.

In the best fits the parameter $g_1$ is found to be varying in the range of $2.12-5.38~\mr{GeV^{-1/2}}$ which is
compatible with that determined in Ref.~\cite{Colangelo:2002mj}. Notice that this quantity still has large
uncertainties as pointed out in Ref.~\cite{Colangelo:2002mj}. We still regard the fitted values as reasonable.
All the fitted parameters are listed in Table~\ref{tab:para} for the SU(2) and SU(3) scheme, respectively.

\begin{table*}[t]
\begin{center}
\renewcommand{\arraystretch}{1.3}
\caption{Fitted parameters in schemes with SU(2) and SU(3) flavor symmetry for the light quark sector.
}\label{tab:para}
\begin{tabular}{c|c|c|cccc}\hline\hline
    & $\Lambda~(\mr{GeV})$&$\chi^2$&$ g_{SP\gamma}~(\mr{GeV^{-1}})$ & $g_{DP\gamma}~(\mr{GeV^{-1}})$ &$g_{1}~(\mr{GeV^{-1/2}})$&$ \delta$
    ($^{\circ}$)\\\hline
&0.8&1.95&0.232$\pm$0.026&0.314$\pm$0.056&2.12$\pm$1.43&$-119\pm$11\\
&0.9&1.56&0.222$\pm$0.029&0.324$\pm$0.029&4.44$\pm$1.71&$-114\pm$14\\
SU(2)&1.0&1.20&0.214$\pm$0.055&0.331$\pm$0.025&5.37$\pm$0.51&$-109\pm$14 \\
&1.1&0.94&0.208$\pm$0.048&0.335$\pm$0.022&5.31$\pm$1.61&$-106\pm $12\\
&1.2&0.78&0.206$\pm$0.048&0.336$\pm$0.027&4.49$\pm$1.02&$-105\pm$13\\\hline
&0.8&1.61&0.219$\pm$0.030&0.326$\pm$0.026&4.52$\pm$0.72&$-114\pm$10\\
&0.9&1.16&0.209$\pm$0.049&0.332$\pm$0.022&5.38$\pm$1.17&$-106\pm$9\\
SU(3)&1.0&0.88&0.205$\pm$0.047&0.335$\pm$0.021&4.70$\pm$1.22&$-104\pm$11\\
&1.1&0.74&0.207$\pm$0.048&0.335$\pm$0.026&3.38$\pm$1.29&$-104\pm$14\\
&1.2&0.66&0.209$\pm$0.049&0.334$\pm$0.028&2.49$\pm$0.64&$-104\pm$14
\\\hline\hline%
\end{tabular}
\end{center}
\end{table*}

\begin{table*}
\centering \caption{The full calculation results for the partial decay widths in the SU(2) and SU(3) scheme are
listed to compare with the quark model calculations and experimental data. All values are given in keV except
that the values for $\Lambda$ is given in GeV.  The NR and relativized Godfrey-Isgur model results updated in
Ref.~\cite{Barnes:2005pb} are quoted as (a) and (b), respectively. The long-dashed line ``---" denotes
unavailability of the corresponding model calculation or experimental data from BESIII. } \vskip 0.3cm
\begin{ruledtabular}
\renewcommand{\arraystretch}{1.3}
\begin{tabular}{c|cccc|ccccc|ccccc|cc}
Process &\multicolumn{4}{c|}{Quark Model} & \multicolumn{5}{c|}{SU(2) with $\Lambda=$}
&\multicolumn{5}{c|}{SU(3) with $\Lambda=$}&\multicolumn{2}{c}{Experiment }
\\ & \cite{Eichten:1979ms} & \cite{Brambilla:2004wf} & \cite{Barnes:2005pb}(a) & \cite{Barnes:2005pb}(b)
&0.8&0.9&1.0&1.1&1.2&0.8&0.9&1.0&1.1&1.2&  PDG~\cite{Agashe:2014kda} &  BESIII~\cite{Ablikim:2015sol,BESIII:2015cby}
\\\hline
 $\psi''\to\chi_{c0}\gamma$ &---  &299 &403&213&199&199&197 &196 &194 &199 &197 &195 &194&193 &$199\pm26$ &$187\pm20$\\
 $\psi''\to\chi_{c1}\gamma$ &---  & 99  &125& 77&64.5&64.8&65.6&66.6&67.4&64.8&66.0&67.2&67.8&68.0&$73.4\pm13.9$ & $67.46\pm7.85$  \\
 $\psi''\to\chi_{c2}\gamma$&---   &3.88& 4.9& 3.3&3.2  &3.1 & 2.9 &2.9  &3.1 &2.7  &2.5  &2.6  &3.1 &3.5  & $<$24.5 & $<$17.4         \\
 \hline
 $\psi'\to\chi_{c0}\gamma$ & 50  & 47  & 63 & 26&30.3&30.2&30.1&30.0&29.9&30.1&30.0&29.9&29.9&29.9& $29.8\pm1.1$  &--- \\
 $\psi'\to\chi_{c1}\gamma$ &45.3&42.8 & 54 & 29&27.8&28.0&28.2&28.3&28.4&28.0&28.2&28.4&28.4&28.4& $28.5\pm1.2$  &--- \\
 $\psi'\to\chi_{c2}\gamma$ &28.9& 30.1& 38 & 24&19.0&19.0&19.1&19.3&19.4&19.0&19.2&19.3&19.4&19.5& $27.2\pm1.2$&---
\end{tabular}
\end{ruledtabular}
\label{tab:results}
\end{table*}

\begin{table*}
\caption{Individual contributions from different diagrams with $\Lambda=1.0~\mathrm{GeV}$. All values are in unit
of keV. For the SU(3) scheme the mixing and meson loop contributions are presented  with (left column) and
without (right column) contributions from the charmed-strange meson loops after the parameters are fitted. }
\vskip 0.3cm
\begin{ruledtabular}
\renewcommand{\arraystretch}{1.3}
\begin{tabular}{c|ccccc|ccccccccccccc}
Process &\multicolumn{5}{c|}{SU(2)} & \multicolumn{13}{c}{SU(3)}
\\ &Tree & Mixing & Triangle &Gauging& Full & Tree &&  \multicolumn{2}{c}{Mixing}  && \multicolumn{2}{c}{Triangle} && \multicolumn{2}{c}
{Gauging}&& \multicolumn{2}{c} {Full}
\\\hline
 $\psi''\to\chi_{c0}\gamma$ &232 & 1.16   & 2.79 &4.87  &197 &238&&1.19   &1.06   && 3.47 &2.12   && 6.18&3.70 && 195&210  \\
 $\psi''\to\chi_{c1}\gamma$ &73.2& 1.46   & 0.73 &0.62  &65.6&75.1&&1.50   &1.34   && 0.89 &0.55  && 1.11 &0.47 &&67.2&71.7 \\
 $\psi''\to\chi_{c2}\gamma$ &2.8  & 1.42   &0.024&0.013&2.9  & 2.9 &&1.46   &1.30   &&0.028&0.018&&0.027&0.010&&2.6 &2.7    \\  \hline
 $\psi'\to\chi_{c0}\gamma$  &25.4& 0.17   & 0.34 &0.31  &30.1&23.4&&0.23   &0.18   && 0.35 &0.18  &&0.53 &0.24  &&29.9&27.4  \\
 $\psi'\to\chi_{c1}\gamma$  &22.1& 0.037 & 0.22 &0.26  &28.2&20.3&&0.050 &0.038  && 0.40 &0.17  &&0.52 &0.20  &&28.4& 25.1 \\
 $\psi'\to\chi_{c2}\gamma$  &15.4&0.0010& 0.12 &0.16  &19.1&14.2&&0.0014&0.0011&& 0.24&0.090 && 0.33 &0.12  &&19.3&17.1
\end{tabular}
\end{ruledtabular}
\label{tab:exp}
\end{table*}

The best fitted partial widths in the SU(2) and SU(3) schemes are listed in Table~\ref{tab:results}. The results
are compared to three particular model calculations, i.e. Cornell model~\cite{Eichten:1979ms}, non-relativistic
quark model (NR), and relativized quark model~\cite{Barnes:2005pb}, as well as to experimental data from
PDG~\cite{Agashe:2014kda,BESIII:2015cby} and recent measurement by BESIII~\cite{Ablikim:2015sol}. It shows that
the non-relativistic quark model gives rather large partial widths for the E1 transitions $\psi''\to
\gamma\chi_{cJ}$, nearly twice the experimental values. The Cornell model also over-shoots the data quite
significantly while only the relativized Godfrey-Isgur model appears to have some agreement with the data. Our
model can fit well these five channels, i.e. $\psi''\rar\gamma\chi_{c0,1,2}$ and $\psi'\rar\gamma\chi_{c0,1}$,
in both SU(2) and SU(3) scheme which hints an insignificant role played by the charmed-strange meson loops.  For
the channel of $\psi'\rar\gamma\chi_{c2}$, as mentioned before, we take our result as a prediction for future
measurement which will be more precise.

To better understand our results we compare the exclusive contributions from those different transition
processes in Fig.~\ref{fig:nlo} with $\Lambda=1.0\mathrm{GeV}$. Thus, we list the partial widths from those
exclusive processes in Table~\ref{tab:exp} for both SU(2) and SU(3) scheme, respectively, and once again the
results for the full calculations are shown as a comparison. For both schemes the tree-level transitions are
dominant as expected and in channels of $\psi''\to \gamma\chi_{cJ}$ they can be compared with the
non-relativistic quark model calculations since the loops give destructive inferences. In channels of $\psi'\to
\gamma\chi_{cJ}$ they are smaller compared with the non-relativistic quark model calculations since the loops
give constructive inferences while the quark model calculations already give larger results than the
experimental data. In the NREFT scenario the effective coupling determined at tree level can be regarded
equivalent to the combined coupling strength from the wave function overlap and spin-flavor factors in the
non-relativistic quark model. But it should be noted that in our NREFT formulation the mixing and triangle loop
processes serve as important corrections to the leading non-relativistic results which is different from the
relativistic corrections introduced in the relativized quark model~\cite{Barnes:2005pb}. In the latter the
relativistic corrections are considered by the Lorentz boost factor for the constituent quarks. Here,  the
corrections arise from the intermediate charmed meson degrees of freedom which suggests that virtual states
involving creations of light quark pairs from vacuum are essential. Especially, such a mechanism becomes
important when the threshold of the intermediate mesons is close to the mass of the coupled state.

From Table~\ref{tab:exp} one can see that the triangle loop and the EM gauging process have larger corrections
than the mixing process. This is a further indication for the necessity of including the meson loop transitions
as leading corrections to the E1 transitions. The correction from the mixing terms can be regarded as kind of
wave function corrections if we compare it with the quark model approach. But the triangle loop and the EM
gauging term represent a different mechanism compared to the quark model picture and mimic the unquenched
effects that have not been included in the constituent quark model. An interesting consequence is that the
static properties of both $\psi'$ and $\psi''$ will not be affected significantly by the meson loops, for
instance, their masses and total widths, etc. However, their decay modes can recognize the effects arising from
the loop transitions. This explains the success of the non-relativistic quark model in the description of the
charmonium spectrum near the $D\bar{D}$ threshold.  But significant discrepancies between the theoretical
calculations and experimental data were found even for the E1 transition calculations.

We should mention that our calculations give better results than non-relativistic quark models in most channels
except for $\psi'\to\gamma\chi_{c2}$. Note that the PDG averaged value for $\psi'\to\gamma\chi_{c2}$ is based on
several measurements~~\cite{Agashe:2014kda} among which significant discrepancies can be seen. This may explain
that it is hard to accommodate the $\psi'\to\gamma\chi_{c2}$ channel in the numerical fitting. We anticipate
that more precise measurement of $\psi'\to\gamma\chi_{cJ}$ will be able to provide more stringent constraints on
our model parameters and also examine our scenario in the future.

In Table~\ref{tab:exp} the SU(3) scheme is presented with (left column) and without (right column) contributions
from the charmed-strange meson loops with the parameters fixed in the fitting. Although the charmed-strange
meson loop  can bring some changes to the parameters listed in Table~\ref{tab:para}, it is consistent to be
small due to the relatively larger mass of the $D_s \bar{D}_s^*+c.c.$ threshold.

%
\begin{figure}[t] \vspace{0.cm}
\begin{center}
\includegraphics[scale=0.5]{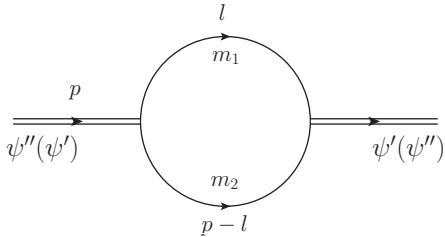}
\caption{The mixing diagram for $\psi''(\psi')\to \psi'(\psi'')$ via intermediate $D$-meson loops. }
\label{fig:loopint}
\end{center}
\end{figure}

We can further investigate the mixing term and extract the mixing angle between $\psi'$ and $\psi''$. For the
mixing process of Fig.~\ref{fig:loopint} we first define the mixing parameter
$|\xi|$~\cite{Achasov:1979xc,Wu:2008hx,Zhang:2009gy},
\be
|\xi_i|\equiv\left|\fc{D_{\psi'\psi''}}{D_{i}}\right|,
\ee
where $D_{\psi'\psi''}$ is the mixing term via heavy meson loop and $D_{i}$ ($i=\psi', \ \psi''$) is the
denominator for the propagator of $\psi'$ or $\psi''$. Both of them depend on the initial energy $s$. In our
calculation, it is convenient to extract $|\xi|$ by just taking the ratio of the amplitude of the mixing diagram
Fig.~\ref{fig:nlo}(b) for $\psi''(\psi')$ and the amplitude of  the tree diagram Fig.~\ref{fig:nlo}(a) for
$\psi'(\psi'')$:
\be
|\xi_i|\equiv\lt|\fc{D_{\psi'\psi''}}{D_{i}}\rt|=\lt|\fc{\Amp^{mixing}_{\psi''(\psi')}}{\Amp^{tree}_{\psi'(\psi'')}}\rt|.
\ee
So the mixing parameter $|\xi_{\psi'}(s)|$ at $\st s=m_{\psi''}=3.773~\mr{GeV}$ can be related to the
$\psi(2S)-\psi(1D)$ state mixing angle via
$|\xi_{\psi'}(s)|\approx|\mr{sin}\theta_{\psi'}|$~\cite{Rosner:2004wy} in which we find
$\theta_{\psi'}\approx8.1^{\circ}$ in SU(2) scheme. This value is consistent with those extracted in
Refs.~~\cite{Rosner:2004wy,Zhang:2009gy}. In the same way we find $\theta_{\psi''}\approx 2.4^{\circ}$ at $\sqrt
s=m_{\psi'}=3.686~\mr{GeV}$, which should be useful for further studies of issues related to the ``$\rho$-$\pi$
puzzle" (see Ref.~\cite{Zhao:2010ja} for a recent review of this topic).

\section{Summary}

We present a detailed study of the E1 transitions for $\psi'$ and $\psi''\to \gamma\chi_{cJ}$ in the NREFT where
the subleading corrections arising from the charmed meson loops are consistently taken into account in the same
framework.
We find that the intermediate meson loops play an important role by introducing destructive interferences that
in most channels bring down the leading contributions from the tree-level transitions. This special mechanism
actually accounts for unquenched effects that have not been included in the constituent quark model. We
emphasize that the open charm contribution from the triangle processes appears to be a general phenomenon that
has brought a lot of interesting insights into the understanding of recent ``XYZ" states. Meanwhile, as we have
shown in this work, it can also produce sizable effects on processes where the dominant contribution is from the
potential quark model. We expect that further precise measurement of $\psi''\to \gamma\chi_{c2}$ at BESIII and
Belle-II will help clarify the underlying dynamics.

\section*{Acknowledgments}
%

This work is supported, in part, by the National Natural Science Foundation of China (Grant Nos. 11425525 and
11521505), DFG and NSFC funds to the Sino-German CRC 110 ``Symmetries and the Emergence of Structure in QCD''
(NSFC Grant No. 11261130311), National Key Basic Research Program of China under Contract No. 2015CB856700. M.C.
is also supported by the Chinese Academy of Sciences President's International Fellowship Initiative Grant
2015PM006.

\begin{appendix}
%
\section{Integrals}\label{app:mixing}

We calculated different kinds of loop integrals with the exponential form factor in Eq.~(\ref{eq:ff}). Two kinds
of two-point loop integrals in the mixing and EM gauging terms are defined as:
\ba
I_{m}(m_1,m_2,M_,\Lam)&=&\intlf\fc{\vec l^2\mr{exp}(-2\vec l^2/\Lam^2)}{(l^2-m_1^2+i\eps)[(P-l)^2-m_2^2+i\eps]},\\
I_{g}(m_1,m_2,M_,\Lam)&=&\intlf\fc{\mr{exp}(-2\vec l^2/\Lam^2)}{(l^2-m_1^2+i\eps)[(P-l)^2-m_2^2+i\eps]}.
\ea
Three kinds of triangle loop integrals are defined as:
\ba
I^{(0)}(m_1,m_2,m_3,M,M_1,M_2,\Lam)&=&i\intlf\fc{\mr{exp}(-2\vec l^2/\Lam^2)}{(l^2-m_1^2+i\eps)[(P-l)^2-m_2^2+i\eps][(l-q)^2-m_3^2+i\eps]},\\
q^i I^{(1)}(m_1,m_2,m_3,M,M_1,M_2,\Lam)&=&i\intlf\fc{l^i\mr{exp}(-2\vec
l^2/\Lam^2)}{(l^2-m_1^2+i\eps)[(P-l)^2-m_2^2+i\eps][(l-q)^2-m_3^2+i\eps]},\\
q^iq^jI^{(2)}_0(m_1,m_2,m_3,M,M_1,M_2,\Lam)&+&\del^{ij}\vec q^2 I^{(2)}_1(m_1,m_2,m_3,M,M_1,M_2,\Lam)\nn\\
&=&i\intlf\fc{l^il^j\mr{exp}(-2\vec l^2/\Lam^2)}{(l^2-m_1^2+i\eps)[(P-l)^2-m_2^2+i\eps][(l-q)^2-m_3^2+i\eps]},
\ea
where $P=(M,\vec0)$ in the rest frame of the initial particle. Since $q$ is the 4-momentum of the photon in
this work, we can set the 3-momentum direction of it to be along $z$-axis. Then $q=(q_z,0,0,q_z)$ and
\be
q_z=\fc{\st{[M^2-(M_1+M_2)^2][M^2-(M_1-M_2)^2]}}{2M}.
\ee

For $I_{m}$ and $I_{g}$ it is straightforward:
\ba
I_{m}&=&\intlf\fc{\vec l^2\mr{exp}(-2\vec l^2/\Lam^2)}{(l^2-m_1^2+i\eps)[(P-l)^2-m_2^2+i\eps]}\nn\\
&=&\fc{i}{4m_1m_2}\intlt\fc{\vec l^2\mr{exp}(-2\vec l^2/\Lam^2)}{P-m_1-m_2-\vec l^2/2\mu_{12}}\nn\\
&=&\fc{i}{4m_1m_2}\fc{4\pi}{(2\pi)^3}\intlz\fc{l^4\ffn}{P-m_1-m_2-l^2/2\mu_{12}}\nn\\
&=&\fc{-i\expklm}{8(m_1+m_2)\pi^2}\lt\{\st{\fc{\pi}{2}}\Lam
e^{2k^2/\Lam^2}\fc{\Lam^2+4k^2}{4}+\pi(-k^2-i\eps)^{3/2}\lt[1-\mr{erf}\lt(\fc{\st{-2k^2-i\eps}}{\Lam}\rt)\rt]\rt\},\\
I_{g}&=&\intlf\fc{\mr{exp}(-2\vec l^2/\Lam^2)}{(l^2-m_1^2+i\eps)[(P-l)^2-m_2^2+i\eps]}\nn\\
&=&\fc{i}{4m_1m_2}\fc{4\pi}{(2\pi)^3}\intlz\fc{l^2\ffn}{P-m_1-m_2-l^2/2\mu_{12}}\nn\\
&=&\fc{i}{4m_1m_2}\lt\{-\fc{\mu\Lam}{(2\pi)^{3/2}}+\fc{\mu
k}{2\pi}e^{-2k^2/\Lam^2}\lt[\mr{erfi}\lt(\fc{\st2k}{\Lam}\rt)-i\rt]\rt\},
\ea
where $k=\st{2\mu(M-m_1-m_2)}$ and
$\mu_{ij}$ is the reduced mass of the intermediate particles which are labeled as $i$ and $j$. The error function
and the imaginary error function are defined as
\ba
\mathrm{erf}(z)&=&\fc{2}{\st\pi}\int^z_0\!e^{-t^2}\mr{d}t,\\
\mathrm{erfi}(z)&=&\fc{2}{\st\pi}\int^z_0\!e^{t^2}\mr{d}t.
\ea
For the triangle integral, we transform them to
be the integral of the Feynman parameter $x$ then do the numerical integrals of $x$. For example:
\ba
I^{(0)}&=&i\intlf\fc{\mr{exp}(-2\vec l^2/\Lam^2)}{(l^2-m_1^2+i\eps)[(P-l)^2-m_2^2+i\eps][(l-q)^2-m_3^2+i\eps]}\nn\\
&=&\fc{\mu_{12}\mu_{23}}{2m_1m_2m_3}\intx\intlt\fc{\ff}{(\vec l^2+\Del)^2}\nn\\
&=&\fc{-\mu_{12}\mu_{23}}{16m_1m_2m_3\Lam^2\pi^2}\intx\lt\{2\Lam\st{2\pi}+\pi
e^{2\Del/\Lam^2}\lef4\st\Del+\fc{\Lam^2}{\st\Del}\rig\lt[\mr{erf}\lef\fc{\st{2\Del}}{\Lam}\rig-1\rt]\rt\},
\ea
where $\Del=x\lef c'-ax\rig+\lef 1-x\rig\lef c-i\eps\rig$. Here $c', a$ and $c$ are defined as in
Ref.~\cite{Guo:2010ak}. With the same method we have
\ba
I^{(1)}&=&\fc{-\mu_{12}\mu_{23}^2}{16m_1m_2m_3^2\pi^2\Lam^2}\intx x\lt\{2\Lam\st{2\pi}+\pi
e^{2\Del/\Lam^2}\lef4\st\Del+\frac{\Lam^2}{\st\Del}\rig\lt[\mr{erf}\lef\fc{\st{2\Del}}{\Lam}\rig-1\rt]\rt\},\\
I^{(2)}_1&=&\fc{\mu_{12}\mu_{23}}{48m_1m_2m_3
q_z^2\pi^2\Lam^2}\intx\lt\{\Lam\st{2\pi}\lef2\Del+\Lam^2\rig+\pi\st\Del
e^{2\Del/\Lam^2}\lef4\Del+3\Lam^2\rig\lt[\mr{erf}\lef\fc{\st{2\Del}}{\Lam}\rig-1\rt]\rt\}.
\ea


\section{Decay amplitudes}\label{app:amplitudes}
%
Using the Lagrangians introduced in Section II, the amplitudes for all mixing diagrams can be expressed as:
\ba\label{mixing-complete-form}
\Amp_{\psi''\rar\gam\chi_{c0}}^{mixing}&=&-\fr{8\st5}{9}g_{2}g_{3}g_{SP\gam}E_{\gam}\eps^{i}_{\psi''}\eps^{*}_{\gam
i}I_m(D^{(*)},\bar D^{(*)})/(m_{\psi''}^2-m_{\psi'}^2+im_{\psi'}\Gam_{\psi'})+c.c.,
\\\Amp_{\psi''\rar\gam\chi_{c1}}^{mixing}&=&\fr{4}{3}\st{\fr{10}{3}}g_{2}g_{3}g_{SP\gam}E_{\gam}\eps_{ijk}\eps^{i}_{\psi''}\eps^{*j}_{\chi_{c1}}\eps^{*k}_{\gam}/(m_{\psi''}^2-m_{\psi'}^2+im_{\psi'}\Gam_{\psi'})+c.c.,
\\\Amp_{\psi''\rar\gam\chi_{c2}}^{mixing}&=&\fr{8}{3}\st{\fr{5}{3}}g_{2}g_{3}g_{SP\gam}E_{\gam}\eps^{i}_{\psi''}\eps^{*j}_{\gam}\eps_{*\chi_{c2}ij}I_m(D^{(*)},\bar
D^{(*)})/(m_{\psi''}^2-m_{\psi'}^2+im_{\psi'}\Gam_{\psi'})+c.c.,
\\\Amp_{\psi'\rar\gam\chi_{c0}}^{mixing}&=&-\fr{40}{9\st3}g_{2}g_{3}g_{DP\gam}E_{\gam}\eps^{i}_{\psi'}\eps^{*}_{\gam i}I_m(D^{(*)},\bar
D^{(*)})/(m_{\psi'}^2-m_{\psi''}^2+im_{\psi''}\Gam_{\psi''})+c.c.,
\\\Amp_{\psi'\rar\gam\chi_{c1}}^{mixing}&=&-\fr{10\sqrt2}{9}g_{2}g_{3}g_{DP\gam}E_{\gam}\eps_{ijk}\eps^{i}_{\psi'}\eps^{*j}_{\chi_{c1}}\eps^{*k}_{\gam}I_m(D^{(*)},\bar
D^{(*)})/(m_{\psi'}^2-m_{\psi''}^2+im_{\psi''}\Gam_{\psi''})+c.c.,
\\\Amp_{\psi'\rar\gam\chi_{c2}}^{mixing}&=&\fr{4}{9}g_{2}g_{3}g_{DP\gam}E_{\gam}\eps^{i}_{\psi'}\eps^{*j}_{\gam}\eps_{*\chi_{c2}ij}I_m(D^{(*)},\bar
D^{(*)})/(m_{\psi'}^2-m_{\psi''}^2+im_{\psi''}\Gam_{\psi''})+c.c.,
\ea
where $I_m(D^{(*)},\bar D^{(*)})$ is the
sum of integrals for all possible intermediate $D$-meson loops with appropriate  incoming and outgoing particle
corresponding to the specific channel. For example, $I_m(D^{(*)},\bar D^{(*)})=I_m(m_{D^{(*)}},m_{\bar
D^{(*)}},m_{\psi''},\Lam)$ in Eq.~(\ref{mixing-complete-form}). The gauging amplitudes are:
\ba
\Amp_{\psi''\rar\gam\chi_{c0}}^{gauging}&=&\fr{\st5}{3}ieg_1g_3\epstz\lt[3I_g(\dpl,\dm)+I_g(\dsp,\dsm)\rt]+c.c.,\\
\Amp_{\psi''\rar\gam\chi_{c1}}^{gauging}&=&2\st{\fr{5}{6}}ieg_1g_3\epsto I_g(\dpl,\dsm)+c.c.,\\
\Amp_{\psi''\rar\gam\chi_{c2}}^{gauging}&=&-\fr{i}{\st{15}}eg_1g_3\epstt I_g(\dsp,\dsm)+c.c.,\\
\Amp_{\psi'\rar\gam\chi_{c0}}^{gauging}&=&\fr{i}{\st3}eg_1g_2\epstz\lt[3I_g(\dpl,\dm)+I_g(\dsp,\dsm)\rt]+c.c.,\\
\Amp_{\psi'\rar\gam\chi_{c1}}^{gauging}&=&-2\st2ieg_1g_2\epsto I_g(\dpl,\dsm)+c.c.,\\
\Amp_{\psi'\rar\gam\chi_{c2}}^{gauging}&=&-2ieg_1g_2\epstt I_g(\dsp,\dsm)+c.c.,
\ea
where $e$ is the unit
charge.

\begin{table*}[t]
\begin{center}
\caption{Intermediate charmed meson loops contributing to each transition. The loops are denoted as $[m_1,m_2,m_3]$
for $m_1$ and $m_2$ rescattering into final $\gamma\chi_{cJ}$ by exchanging $m_3$. The charge conjugation terms are
dropped for simplicity.}\label{tab:ls}
\renewcommand{\arraystretch}{1.3}
\begin{tabular}{c|c|c}\hline\hline
Channels                         &Electric&Magnetic\\\hline
$\psi'(\psi'')\rar\gam\chi_{c0}$ &$[\dpl,\dm,\dpl],[\dsp,\dsm,\dsp]$& $[D^*,\bar D, D]$, \ $[\bar D,D^*,\bar D^*]$ \\
$\psi'(\psi'')\rar\gam\chi_{c1}$ &$[\dpl,\dsm,\dpl],[\dsp,\dm,\dsp]$& $[D,\bar D,D^*]$, \ $[D^*,\bar D^*,D]$, \ $[D^*,\bar D,D^*]$ \\
$\psi'(\psi'')\rar\gam\chi_{c1}$ &$[\dsp,\dsm,\dsp]$& $[\bar D,D^*,\bar D^*]$, \ $[D^*,\bar D^*,D^*]$
\\\hline\hline%
\end{tabular}
\end{center}
\end{table*}

The amplitudes for all the triangle loop diagrams include electric and magnetic ones due to the different photon
coupling. In Table~\ref{tab:ls} the contributing loops are denoted by the rescattering and exchanging mesons
for each decay channel.
\begin{eqnarray}
\Amp^{triangle}_{\psi''\rar\gam\chi_{c0}}&=&ig_1g_3\epstz q^2\lt[\sq5F_{pv}\inr(D^*,\bar
D,D)+\fr{\sq5}{3}F_{pv}\inr(\bar D,D^*,\bar D^*)\rt]\nn
\\&+&g_3g_1\epstz q_z^2\lt[\fr{4\st5e}{m_D}\ito(\dpl,\dm,\dpl)+\fr{4\st5e}{3m_{D^*}}\ito(\dsp,\dsm,\dsp)\rt]+c.c.,
\\\Amp^{triangle}_{\psi''\rar\gam\chi_{c1}}&=&-2\sq{\fr{10}{3}}ig_1g_3\eps^i_\psi\eps^{*j}_\chi\eps^{*l}_\gam q_iq^k\eps_{jkl}F_{pv}\inr(D,\bar
D,D^*)\nn
\\&-&\sq{\fr{2}{15}}ig_1g_3\eps^i_\psi\eps^{*j}_\chi\eps^{*l}_{\gam}q^k\lef4q_i\eps_{jkl}-q_j\eps_{ikl}\rig F_{pv}\inr(D^*,\bar D^*,D)\nn
\\&-&\sq{\fr{10}{3}}ig_1g_3\eps^i_\psi\eps^{*j}_\chi\eps^{*l}_\gam q_jq^k\eps_{ikl}F_{vv}\inr(D^*,\bar D,D^*)\nn
\\&+&2\st{\fr{10}{3}}eg_1g_3\epsto q_z^2\lt[\ito(\dpl,\dsm,\dpl)/m_D+\ito(\dsp,\dm,\dsp)/m_{D^*}\rt]+c.c.,
\\\Amp^{triangle}_{\psi''\rar\gam\chi_{c2}}&=&-\sq{\fr{5}{3}}ig_1g_3\eps^i_\psi\eps^{*n}_{\gam}q^jq^m\eps_{ijk}\eps_{lmn}\lef\eps^{*kl}+\eps^{*lk}\rig
F_{pv}\inr(\bar D,D^*,\bar D^*)\nn
\\&-&\fr{1}{\sq{15}}ig_1g_3\lef\eps_{\psi i}q^j+\eps^j_\psi q_i\rig\lef\eps^{*}_{\gam j}q_k-\eps^{*}_{\gam
k}q_j\rig\lef\eps^{*ik}_\chi+\eps^{*ki}_\chi\rig F_{vv}\inr(D^*,\bar D^*,D^*)\nn
\\&-&\fr{4eg_1g_3}{\st{15}m_{D^*}}\lef\del_{ik}\del_{jl}+\del_{il}\del_{jk}\rig\eps^i_\psi\eps^{*j}_{\gam}\eps^{*kl}_\chi
q_z^2\ito(\dsp,\dsm,\dsp)+c.c.,
\\\Amp^{triangle}_{\psi'\rar\gam\chi_{c0}}&=&-ig_1g_2\epstz q_z^2\lt[2\sq3F_{pv}\inr(D^*,\bar D,D)+\fr{2}{\sq3}F_{pv}\inr(\bar D,D^*,\bar
D^*)\rt]\nn
\\&+&g_1g_2\epstz q_z^2\lt[\fr{4\st3e}{m_D}\ito(\dpl,\dm,\dpl)+\fr{4e}{\st3m_{D^*}}\ito(\dsp,\dsm,\dsp)\rt]+c.c.,
\\\Amp^{triangle}_{\psi'\rar\gam\chi_{c1}}&=&-2\sq2ig_1g_2\eps^i_\psi\eps^{*j}_\chi\eps^{*l}_\gam q_iq^k\eps_{jkl}\lt[F_{pv}\inr(D,\bar
D,D^*)+F_{pv}\inr(D^*,\bar D^*,D)\rt]\nn
\\&+&2\sq2ig_1g_2\eps^i_\psi\eps^{*j}_\chi\eps^{*l}_\gam q_jq^k\eps_{ikl}\lt[F_{pv}\inr(D^*,\bar D^*,D)+F_{vv}\inr(D^*,\bar D,D^*)\rt]\nn
\\&-&2\st2eg_1g_2\epsto q_z^2\lt[\ito(\dpl,\dsm,\dpl)/m_D+\ito(\dsp,\dm,\dsp)/m_{D^*}\rt]+c.c.,
\\\Amp^{triangle}_{\psi''\rar\gam\chi_{c2}}&=&2ig_1g_2\eps^i_\psi\eps^{*n}_{\gam}q^jq^m\eps_{ijk}\eps_{lmn}(\eps^{*kl}_\chi+\eps^{*lk}_\chi)F_{pv}\inr(\bar
D,D^*,\bar D^*)\nn
\\&-&2ig_1g_2\lef\eps_{\psi i}q^j+\eps^j_\psi q_i\rig\lef\eps^{*}_{\gam j}q_k-\eps^{*}_{\gam k}q_j\rig\lef\eps^{*ik}_\chi+\eps^{*ki}_\chi\rig
F_{vv}\inr(D^*,\bar D^*,D^*)\nn
\\&-&\fr{8eg_1g_2}{m_{D^*}}\lef\del_{ik}\del_{jl}+\del_{il}\del_{jk}\rig\eps^i_\psi\eps^{*j}_{\gam}\eps^{*kl}_\chi
q_z^2\ito(\dsp,\dsm,\dsp)+c.c.,
\end{eqnarray}
where $F_{pv}$ and $F_{vv}$ stand for the charge factors arising from the photon and charmed meson coupling
vertices for different channels, i.e. $F_{pv(n)}=2e\beta/3+2e/(3m_Q)$ for neutral pseudoscalar and vector
charmed mesons, $F_{pv(c)}=-e\beta/3+2e/(3m_Q)$ for charged pseudoscalar and vector charmed mesons, $F_{vv
(n)}=2e\beta/3-2e/(3m_Q)$ for neutral vector charmed mesons and $F_{vv(c)}=-e\beta/3-2e/(3m_Q)$ for charged
vector charmed mesons.
\end{appendix}

\end{document}